\documentclass[aps,prb,superscriptaddress,showpacs,preprint]{revtex4}

\usepackage{graphicx}
\usepackage{bm}
\usepackage{amsmath}
\usepackage[cp1251]{inputenc}
\usepackage[english]{babel}

\input epsf
\usepackage[T2A]{fontenc}

\usepackage{latexsym}
\usepackage{amsbsy}
\usepackage{amsmath}
\usepackage{amssymb}
\usepackage{latexsym}
\usepackage{amsbsy}
\usepackage{amsfonts}
\usepackage[mathcal]{eucal}

\begin{document}

\title{Topologically Ordered Phase States:  
from Knots and Braids to Quantum Dimers}

\author{Luigi Martina}
\affiliation{
Dipartimento di Fisica dell'Universit\`a del Salento and
Sezione INFN di Lecce. \\ Via Arnesano, CP. 193 I-73100 Lecce, 
Italy} 

\author{Alexander Protogenov}
\affiliation{Institute of Applied Physics of the RAS, 603950  Nizhny Novgorod, Russia}
\affiliation{Max Planck Institute for the Physics of Complex
Systems, D-01187 Dresden, Germany}

\author{Valery Verbus}
\affiliation{Institute for Physics of Microstructures of the RAS, 
603950 Nizhny Novgorod, Russia} 

\date{\today}

\begin{abstract}
We consider universal statistical properties of systems that are characterized by phase states with macroscopic degeneracy of the ground state. A possible topological order in such systems is described by non-linear discrete equations. We focus on the discrete equations which take place in the case of generalized exclusion principle statistics. 
We show that their exact solutions are quantum dimensions of the 
irreducible representations of certain quantum group. These solutions provide an example of the point where the generalized exclusion principle statistics and braid statistics meet each other. We propose a procedure to construct the quantum dimer models by means of projection of the knotted field configurations
that involved braiding features of one-dimensional topology.   
\end{abstract}

\pacs{71.10.-w, 71.10.Pm, 71.10.Fd}

\maketitle

\newpage

\section{Introduction}

The universal behaviour of low-dimensional strongly correlated
systems at low temperatures is determined to a great extent by the topology of the manifolds, where the ground
state and low-lying excitations are determined. In strongly correlated electron liquids with high degree of degeneracy of the ground state
such a manifold is presented by a collection of strings in the form of an arbitrary tangle of knotted and linked filaments.
The processes of fusion and decay of the strings give rise to
modifications of the tangle and, consequently,
to the universal character of the quantum criticality in phase states with topological order. Being a result of detailed study of electron liquids in the states with fractional Hall effect, this
conclusion was supported recently by the results, obtained during studying dynamics of spin \cite{SVB,FNS,MSF} and charge \cite{PBSF,FS} degrees of freedom in other low-dimensional electron systems.

It is well known that statistics of excitations in $(1+1)$- and $(2+1)$-dimensional systems is connected with the braid group. Quantum states in such systems are classified by irreducible representations of the braid group instead of the even (for bosons) or odd (for fermions) irreducible representations of the permutation group 
as (3+1)-dimensional systems. One-dimensional irreducible representations of the braid group correspond to Abelian anyon states, while multi-dimensional irreducible representations describe the non-Abelian states of anyons. Statistics of anyon
excitations is called either braid statistics or fractional statistics, because it leads to the conclusion that there exist particles with a fractional charge and spin. In the long-wavelength limit the description
of non-Abelian anyons is based on the effective action of the topological field theory \cite{Wit}, which contains the Chern-Simons term.
An important particular case is the quantum group $SU(N)_{k}$ case,  where the integer $k$ is the coefficient of the Chern-Simons action, or a level in the Wess-Zumino-Witten-Novikov theory.
    
We can also employ another approach \cite{H}, based on the generalized exclusion principle. Studies of the distribution function using this approach \cite{Wu,BW,Nay,ISAK,FK,GS} have shown that this method is equivalent to the thermodynamic Bethe ansatz \cite{Zamo}. The equation, which determines the 
minimum of  free energy, has the form of the Hirota equation \cite{Wieg}. 
It is well known from the theory of nonlinear equations, that this discrete equation in the continuous limit \cite{KWZ} yields the known
integrable hierarchies of nonlinear equations. An important feature
of the derivation of discrete equations in the theory with the
generalized exclusion principle is the absence of any reference
to the dimensionality of the space, unlike braid statistics. Formally generalized exclusion principle statistics \cite{H} may take place not only in low-dimensional systems. The reason of the emergence of the solutions, which exist only in low-dimensional systems, is that in the limit of large values of momentum, discrete equations in the universal sector \cite{ProtVer} of exclusion  statistics encode actually the particle fusion rules, which take place in the conformal field theory. Going over to the limit
$k \gg 1$ in discrete equations of motion, the specifity of low-dimensional situation vanishes.
   
The features of considered statistics are summed up by the theory of tensor categories \cite{Tur,BK}. It unifies consistently the processes of braiding and fusion of string manifolds, which are images of quasiparticle world lines. In the case of low-dimensional systems this occurs in the theory of modular tensor categories \cite{Tur,BK}. The theory of symmetric tensor categories \cite{Tur,Wen,LevWen} is 
applicable to the $(3+1)D$ systems. In the last case, restrictions on solutions of equations
of motion are so strong, that all anyon states are excluded, and only bosons ($\alpha=2\pi$) and fermions ($\alpha=\pi$) with the interchange phase $\alpha$ remain out of anyon states.

To solve the problems of the theory of strongly correlated systems on a lattice, it is often convenient to use the theory of the braid group representation or the Temperley-Lieb algebra (TLA) representations \cite{FF} with a special value of the TLA parameter. In the continuous limit, we can also employ the 
effective Chern-Simons action \cite{FNSWW,AFF}. In particular, to classify the hierarchy of phase states with the aid of $(3+0)$-dimensional spinor Ginzburg-Landau functional \cite{BFN}, it is convenient to use the Hopf invariant 
\cite{VK,PV1,Prot}, which is the $(3+0)D$ analog of the Chern-Simons term.

The construction of a lattice model out from the continuous theory (even inheriting 
its essential properties) is evidently an ambiguous procedure. 
Some intuitive insight of how this can be done, based on the mentioned properties, may include the following consideration. It is well known that the
Chern-Simons term in the action of $(2+1)$-dimensional systems encodes  the invariant description of fluctuating Aharonov-Bohm vortices.
The action of the doubled Chern-Simons theory (for example $SU(N)_{k}\times \overline S \overline U(N)_{k}$ \cite{LevWen,FNSWW,AFF}) which refers to the systems, where $T$ and $P$ invariances are not brocken, includes pairs of Aharonov-Bohm vortices with the opposite chiralities. It is natural to suppose that neutral pairs of Aharonov-Bohm vortices in such systems are 
plane slices of a string loop (from the three-dimensional point of view). A pair of Aharonov-Bohm vortices are portions of the loop cutted by a plane. This means on the whole that a small loop with the scale of the order of the lattice constant 
induces a dimerized configuration of currents when it is projected on the plane.  
Such a projected loop, or equivalently the dimer configuration, can be a building block for the formation of self-organized mesoscale structures in the form of nets. The increased interest to quantum dimer distributions \cite{AFF,MoSo,HKMS} is connected not only with the theory of resonance valence bonds or with the support originating from the experience of the exact solvable models \cite{RokKiv}, but it is also motivated by recent results \cite{FNS,FN} in the field of non-Abelian gauge theory.
   
The goal of this paper is two-fold. In the second section we will discuss 
the opportunities, which appear due to mapping essential configurations of 
the (3+0)-dimensional spinor Ginzburg-Landau model into lattice $(2+1)D$ 
dimer configurations. We will show that two-dimensional dimer field
configurations, distributed on two sub-lattices, may be obtained by projecting three-dimensional current configurations of the so-called toroid phase state. 
In the third section we will consider the solutions of nonlinear discrete equations of the thermodynamic Bethe ansatz and will show their relation with the characteristics arising in the approach, based on the use of braid statistics. They are characterised by the quantum dimension, which is obtained as the solution of the mentioned discrete equations. We will give also some arguments in favor of stability of arising 
mappings of string nets, built of golden chains \cite{FTLTKWF}.

\section{Dimer configurations of vortex pairs}

In this section we show how to construct dimer distributions
starting from Hopf links. Indeed, knots and links of field
configurations appear naturally in the long-wavelength description
of $(3+0)D$ systems with spin as follows.
Let us consider a gauged  Ginzburg-Landau model for the charged
two-components order parameter $\Phi = \left(\phi_{1},\phi_{2}\right)^T$ 
given by the free energy functional
\begin{equation}
F=\int d^{3}x\,\biggl[\sum_{\alpha=1}^{2}\frac{1}{2
m_{\alpha}}\left|\left(\hbar
\partial_{k}+i\frac{2e}{c}A_{k}\right)\phi_{\alpha}\right|^2 +
\frac{(rot \bf A )^{2}}{8\pi} + V(\Phi)\biggr] \, , \label{G1}
\end{equation}
with a generic form of the potential $V(\phi_{1},\phi_{2})$
\cite{Prot} and interacting with an internal vector gauge
potential $\bf A $.

The components of the order parameter can be interpreted as the
hopping and pairing amplitudes, respectively, in a $t-J$ 
model \cite{LNW}. The relations between these amplitudes can be
reshaped in terms of the function $\rho$ and a real  unit
3-vector $\bf n$ defined by
\begin{equation} \rho^2 = \sum_{\alpha=1}^2 \frac{\left|\phi_{\alpha}\right| ^2 }{2\, m_{\alpha}
}, \qquad  n^a = \frac{\overline{\phi_{\alpha}} \sigma^a_{\alpha\,
\beta }\phi_{\beta}}{2 \sqrt{m_{\alpha} m_{\beta}}} ,
\end{equation}
 where
${ \sigma^a }, \; a =  1,2,3$ are the Pauli matrices. 
A residual phase-like degree of freedom is encoded into a momentum
contribution to the effective  gauge field 

\begin{equation} 
{\bf c} = {\bf a} - {\bf A} \, , \, \, \, \, \, \, \, 
{\bf a} = \frac{1}{ \rho^2} \sum_{\alpha=1}^2 \frac{\imath}{2
m_{\alpha}} \left[\phi_{\alpha}\nabla\phi_{\alpha}^{*} - c. c.
\right]. \end{equation}

Thus, one can map the model (\ref{G1}) into 
the extended version of the $O(3)$ nonlinear $\sigma$ model \cite{BFN}

$$
  F = F_{\bf n}+F_\rho+F_{\bf c}+F_{\rm int} =
$$
\begin{eqnarray}
 = \int d^{3}x\left[\frac{1}{4}\rho^{2}\left(\partial_{k}{\bf n}
         \right)^{2}+\left(\partial_{k}\rho \right)^{2}+\frac{1}{16}\rho^{2}
         {\bf c}^{2}+\left(F_{ik}-H_{ik}\right)^{2}+V(\rho,n_{1},
         n_{3})\right],
 \label{G3}
 \end{eqnarray}
where, in dimensionless variables, the field strength
$F_{ik}=\partial_{i}c_k-\partial_{k}c_i$ and the Mermin-Ho 
vorticity $H_{ik} = \partial_{i}a_{k}-\partial_{k}a_{i} = {\bf
n}\cdot[\partial_{i}{\bf n}\times\partial_{k} {\bf n}] $ have been
introduced.

Taking into account only homogeneous density states (i.e. $\rho = const $)  free energy is bounded from below by the inequality
\cite{PV1}
\begin{equation}
F_{\bf n} + F_{\bf c} + F_{\rm int} \geqslant
32\pi^{2}\,|Q|^{3/4}\,(1 - |L|/|Q|)^{2} \, , \label{G4}
\end{equation}
given in terms  of the entries of the symmetric matrix \cite{Prot}
\begin{equation}
K_{\alpha \beta} = 
\frac{1}{16\pi^{2}}\int\limits_\mathcal{M} d^{3}x \,
\varepsilon_{ikl}a_{i}^{\alpha}\partial_{k}a_{l}^{\beta} =
 \left( \begin{array}{cc}
 Q & L  \\
 L & Q^{\prime} \quad
 \end{array} \right), {\rm where} \;  a_i^1 \equiv a_i ,\;  a_i^2 \equiv c_i .
\label{G5}
\end{equation}

Taking the boundary condition ${\bf n}(\infty) \to (0, 0, 1)$, one
compactifies  $\mathbb R^{3}\to  S^{3}$ and $Q$   is the degree of
the mapping  ${\bf n} :   S^{3} \to S^{2}$. That is, $Q$ measures
the linking and knotting of the filaments, which are pre-images 
in $S^{3}$ of an arbitrary chosen value ${\bf n}(x,y,z)={\bf n}_0
\in S^{2}$.  General arguments of the homotopic group theory
\cite{ArKhe} ensure that  $\pi_{3}(S^{2})= \mathbb Z$ , i.e. $Q\in
\mathbb Z$, labelling distinct sectors among all configurations of
the field $\bf n$.  For two linked loops one has $Q = 1$, for the
trefoil knot $Q = 6$ and etc. On the other hand, $L$ is the mutual
linking number \cite{ArKhe} of the fields ${\bf c}$ and ${\bf a}$. 
The integer $Q$ is related \cite{Prot} to the coefficient $k$ in the
Chern-Simons action. For instance, the value $Q=1$ for the Hopf
link of two loops is equivalent to $k=2$.

But, differently from $\bf n$, the length of the vector $\bf c$ is
not fixed and it does not belong to a compact manifold, thus
homotopy group techniques say that  $L$ ( and $Q^{'}$) in
(\ref{G5}) is an arbitrary real numbers. Furthermore, for $|L| <
|Q|$, from Eq. (\ref{G4}), a decreasing  of the lower bound of $F$
is obtained with respect to the case ${\bf c}=0$ \cite{VK}. This
effect is  due to the interaction term $- 2 F_{ik}H_{ik}$ in Eq.
(\ref{G3}). Its contribution could be  so important to completely
compensate the energy contributions $F_{\bf c}$ and $F_{\bf n}$.
Such a situation may occur because all contributions in Eq.
(\ref{G3}) are of the same order. Then, one can prove \cite{Prot}
the existence of inhomogeneous states with total energy smaller
than in the Skyrme-Faddeev reduction (i.e. $\bf c = 0$, $\rho =
const$),  if the knot is contained  in a region of typical size
$\sqrt{2} > R \sim 1/\rho $ and the amplitude of the momentum
field $|{\bf c}| \sim c_0$ is bounded by $\alpha / R_0 < c_0 <
1/R_0$, where $R_0$ can be figured out as the "thickness" of the
filamentary structure of $\bf c$, or $\bf n$, and $\alpha = \left(
R_0/R \right)^2 \ll 1$ is the packing parameter. Thus, by
(\ref{G4}) one can evaluate the decrease of energy, by a negative
(condensation energy) contribution $\Delta F \sim \left(64 \pi^2 /
Q^{1/4} \right) \alpha$.

In the optimum case of great values ${\bf c}$ the self-dual state 
with $F_{\bf n}=F_{\bf c}$ may be considered as the ground state 
with $F_{min}=0$. It is characterized by the dense packing of filaments 
in knots \cite{PV2} and by the condition ${\bf a}-{\bf c} = {\bf A} = 0$
when the matrix
\begin{equation}
K_{\alpha\beta} = Q\left( \begin{array}{cc}
 1 & 1  \\
 1 & -1
\end{array} \right) \, .
\label{G6}
\end{equation}

Indeed, let us assume
that the amplitude of density $\rho^2 \sim R^{-1}$ is large enough
and a knot size $R$ is so small that the packing degree of
filaments in the knot \cite{PV2} $\alpha = \xi/R \lesssim 1$. The
correlation lenght $\xi$ having the order of the lattice constant
is determined the filament thickness. The decrease of our tuning
parameter $\alpha$ with the decrease of $\rho^{2}$ is accompanied
by the transition to a state, in which among planar field
projections the field configurations in the form of closed
one-dimensional distributions are most preferable. In accordance
with Ginzburg's proposal, we will call this phase state with the
spontaneous diamagnetism a toroid state \cite{VGK,DT}. The phase
state with zero value of the total magnetic moment $\sum_{i}{\bf
m}_{i}$, in which current order is characterized by a polar vector
${\bf T}$, that changes sign under time reversal, can be
represented by the ordering of toroid moments. The toroid moment
${\bf T}=\frac{1}{2}\sum_{i}\left[{\bf r}_{i} \times {\bf
m}_{i}\right]$ can be determined in the following way
\cite{VGK,DT}
\begin{equation}
\bf a={\rm curl}\,{\rm curl}\,\bf T \, . \label{G7}
\end{equation}
Using this equation and the identity \cite{MH}
\begin{equation}
{({\rm curl}\,{\bf a})}_i=\frac{1}{2}\,\varepsilon_{ikl}\,{\bf n}
\left[\partial_k{\bf n}\times\partial_l{\bf n}\right] \label{G8}
\end{equation}
one can express the toroid moment via the degree of chirality
$\displaystyle{\bf n}\cdot\left[\partial_k{\bf
n}\times\partial_l{\bf n}\right]$ in the form:

\begin{equation}
T_i({\bf r})=\frac{1}{16\pi^2}\int d^3r'\left[
\frac{({\bf r}-{\bf r}^{\,\prime})_k} 
{|{\bf r}-
{\bf r}^{\,\prime}|^3}\int d^3r''\,
\frac{{\bf n}\cdot\left[\partial_k{\bf n}\times\partial_i{\bf n}\right]}
{{|{\bf r}^{\,\prime}-{\bf r}^{\,\prime\prime}|}}\right].
\label{G9}
\end{equation}

We see that toroid moment ${\bf T}$ is given by the Biot-Savart
law as well as by the factor which is determined by the Coulomb
Green function and by the field strenth $H_{ik}$ of the Hopf
invariant density. The vector ${\bf T}$ characterizes the
distribution of the poloidal component of the current on the torus
\cite{DT}. The magnetic flux of this current is confined to the
interior of the torus. If the degree $\displaystyle {\bf
n}\cdot\left[\partial_k{\bf n} \times\partial_i{\bf
n}\right]=\varepsilon_{ki}\,\delta({\bf r})$ of the ${\bf
n}$-field noncollinearity is localized at a point the toroid
moment has the form $T_i({\bf r})= (8\pi)^{-1}\varepsilon_{ki}x_k
/r$. The toroid moment is perpendicular to the plane, where the
central loop of the torus is situated, and belongs to the
perpendicular plane that intersects this loop.

The significant constraint of the scale of toroid configuration
appears from the analysis of stability of such  field
distributions. As it has been shown in Ref. \cite{W}, that these
configurations are stable with respect to $s$-wave perturbations
if the characteristic scale of linked distribution is small
enough, i.e. it is of the order of several lattice constants.
Therefore, to describe topological states we have to use a lattice
theory. Besides, in addition, the main contributions of field
configurations correspond usually to small values of topological
numbers $Q$, $L$. Keeping all this in mind we can conclude that
the Hopf links of loop pairs (see Fig. 1), that after being
projected on the plane, provide dimer distributions of vortex
pairs which are the the most significant configurations.
\begin{figure}
\begin{center}
\includegraphics[width=0.9\linewidth]{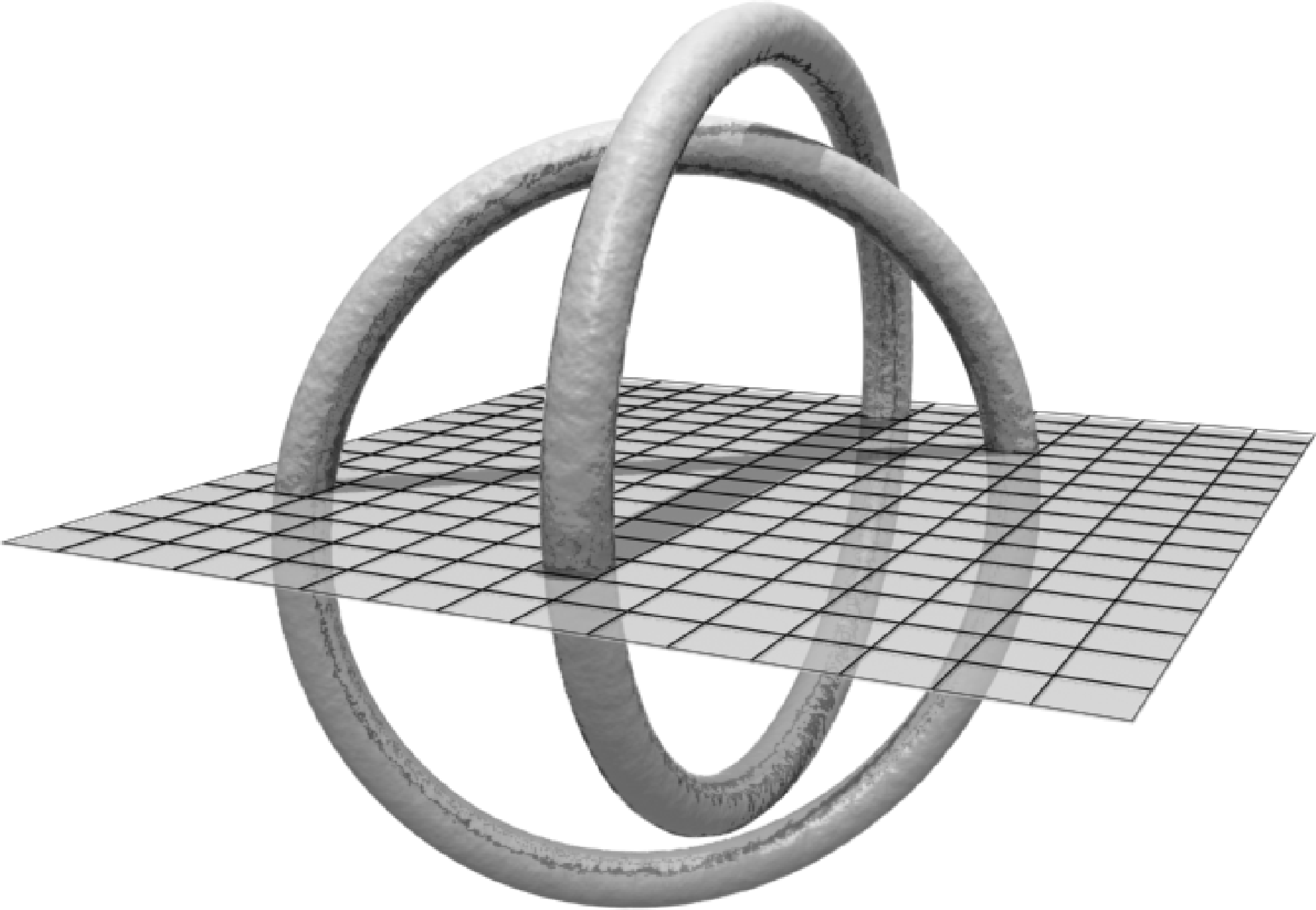}
\end{center}
\caption{(Color online). The Hopf pair with $Q=1$. Their projection into the plane 
presents a system of two dimers, characterized by the respective orientation of 
toroid moments.}
\end{figure}
To deal with states without the time inversion symmetry breaking,
we should take into consideration totally neutral dimer
configurations with antitoroidal ordering of the toroid moments. 

\newpage

\section{Discrete equations of exclusion statistics}
   
Let us consider a system, which contains a set, $\{N_{a}\}$ of
particles with types $a$. The
collective index $a=(\alpha,i)$ contains the index $\alpha$ for
denoting internal degrees of freedom and index $i$ enumerates rapidities of particles. If we fix the variables of all particles,
except the $a$-th one, the $N$-particle wave function can be expressed via the one-particle function of the $a$-th particle.
Let $D_{a}$ be the dimension of such a basis.
Then the rate of changing the number of vacant states due to adding $N_{b}$ particles determines \cite{H} the matrix $g_{ab}$ of statistical  interaction in the following way
\begin{equation}
\frac{\partial D_{a}}{\partial N_{b}}
= -g_{ab}\, .
\label{E1}
\end{equation}

Assuming that the matrix $g_{ab}$ does not depend on the set of numbers $\{N_{a}\}$, we have the solution of the Eq. (\ref{E1}):
\begin{equation}
D_{a} = -\sum_{b} g_{ab}N_{b} + D^{0}_{a}\, .
\label{E2}
\end{equation}
The Eq. (\ref{E2}) contains the number of particles $N_{b}$, added to the system, and the number of vacant states $D^{0}_{a}$
of the $a$-th type in the initial state without particles.
The number of holes $D_{a}$ determines the
statistical weight as follows
\begin{equation}
W = \prod_{a}\frac{(N_{a}+D_{a}-1
+\sum_{b} g_{ab}\delta_{ab})!}
{(N_{a})!
(D_{a}-1+\sum_{b}g_{ab}\delta_{ab})!}\, .
\label{E3}
\end{equation}
In the cases $g_{ab}=0$ and $g_{ab}=
\delta_{ab}$ the Eq. (\ref{E3}) yields well-known statistical weights of Bose and Fermi particles.

The statistical weight $W$ allows to find the entropy $S=\ln W$
and thermodynamical functions.
The free energy in the equilibrium state
\begin{equation}
F=-T\sum_{a}D^{0}_{a}\ln(1 + w_{a}^{-1})
\label{E4}
\end{equation}
is determined by the function $w_{a}$, which can be found from
the equation
\begin{equation}
(1 + w_{a})\prod_{b}\left(1 + w_{b}^{-1}\right)^{-g_{ab}} =
e^{(\epsilon^{0}_{a}-\mu_{a})/T}
\label{E5}
\end{equation}

The variable $w_{a}$ can be expressed via so-called
pseudo-energies $\epsilon=T\ln(D_{a}/N_{a})$ by means of the parametrization $w_{a}=e^{\epsilon_{a}/T}$. In Eq. (\ref{E5}), $T$, $\mu_{a}$ and $\epsilon^{0}_{a}$ are the temperature, the chemical potential and the bare energy of quasiparticles of the type $a$.

We consider the solution of the equation (\ref{E5}) in the limit
$T \gg \epsilon^{0}_{a}-\mu_{a}$. In this case the Eq.
(\ref{E5}) can be written down in the form
\begin{equation}
w_{a} = \prod_{b}\left(1 + w_{b}^{-1}\right)^{N_{ab}}
\label{E6}
\end{equation}
which is typical for the thermodynamic Bethe ansatz.
Here $N_{ab}=g_{ab}-\delta_{ab}$.
 
Below we will be interested in the case of ideal statistics \cite{BW,FK,ProtVer}, when phases of the scattering matrix, being functions of rapidities,
have the structure of step functions. In this case the integral
equation (\ref{E6}) transforms into the algebraic transcendental equation. The matrix $N_{ab}$ can be
expressed \cite{Rav,Ked,Das,Nah} via the incidence matrix
$G_{ab}=\delta_{a+1,b} + \delta_{a,b+1}$ of Lie algebra
with the help of the identity $N=G(2 - G)^{-1}$.
The matrix $2-G$ is the Cartan  matrix of the graph $A_{k+1}/Z_{2}$ \cite{Rav}. Using this identity in Eq. (\ref{E6}) and replacing
$w_{a}=d_{a}^{2}-1$ it is easy to see that the Eq. (\ref{E6}) has
the form
\begin{equation}
d_{a}^{2} = 1 + \prod_{j=1,\, b=2j}^{[k/2]}d_{b}^{G_{ab}} =
\left \{
\begin{array}{ll}
\displaystyle{1+d_2, \; a=1 ,} \\
\displaystyle{1+d_{a-1}d_{a+1}, \; a=2, ...,[k/2]-1,}\\
\displaystyle{1+d_{[k/2]-1}d_{[k/2]}, \; a=[k/2].}
\end{array}
\right.
\label{E7}
\end{equation}
Here index $a$ is connected  with the value of the spin $j$ by the relation $a=2j$, and the upper limit of the product fixes the Jones-Wenzl projector \cite{FF,FNSWW}. 
We will show in Appendix that the Eq. (\ref{E7}) is presented in fact the special 
limit of the Hirota equation \cite{KLWZ}. 

The distribution function 
\begin{equation}
n_{a}=\frac{1}{d_{a}^{2}}=\frac{1}{w_{a}+1}=\frac{1}
{e^{\epsilon_{a}/T}+1}
\label{E8}
\end{equation}
in our case coincides with the probability $p(a{\bar a} \to 0)$ \cite{P} of annihilation of a particle-antiparticle pair in the system of two linked loops of world lines which describe the process of annihilation of
two pairs.

We can find the solution of the Eq. (\ref{E7}) taking into account the appropriate boundary conditions by comparing it with the identity
\begin{equation}
[a]_{q}^{\,2} - 1 = [a+1]_{q}[a-1]_{q} \, .
\label{E12}
\end{equation}
Here $[a]_{q}=(q^{a}-q^{-a})/(q-q^{-1})$, $q=e^{i\pi/(k+2)}$ is the deformation parameter of the $SU(2)_{k}$ Chern-Simons theory.
Identifying $d_{a}=[a+1]_{q}$,
we can see that solutions of the Eq. (\ref{E7})
are quantum dimensions \cite{P,SB,Fen,Aff,K}
\begin{equation}
d_{a}=\frac{\sin[\pi(a+1)/(k+2)]}{\sin[\pi/(k+2)]} \, ,
\label{E13} 
\end{equation}
which are expressed via the Chebyshev polynomials of the second kind,
$U_{m}=\sin[(m+1)\theta]/\sin \theta$ with specification $\theta =
\pi/(k+2)$ for $A_{k+1}$ algebra. In the limit $k \gg 1$, $d_{a=2j}$
equals $2j+1$. The meaning of the quantum dimension is as follows.
The quantum dimension $d_{a}$ determines the rate $d_{a}^{N}$ at which
the dimension of the topological
Hilbert space grows after particles are added.

We pay our attention to the fact that the Eq. (\ref{E7})
is a fermionic  representation \cite{Berc,Ku,Kuni} of the
recursion relation for the Chebyshev polynomials of the second kind. The bosonic representation of recursion relations has the
form $U_{m+1}(x)+U_{m-1}(x)-2xU_{m}(x)=0$. 
From this point of view, we can call the Eq. (\ref{E11}) (see Appendix) the anyon representation of recursion relations for the Chebyshev polynomials of the second kind.  
The roots of the Chebyshev polynomials, being the eigen
values of the matrix $G$, are equal to
\begin{equation}
x_{m,k}=q^{m+1}+q^{-(m+1)}=2\cos \left(\frac{(m+1)\pi}{k+2}\right) \, .
\label{E14}
\end{equation}
The greatest eigen value $x_{0,k}$ of the incidence matrix $G$
is given by the Baraha numbers 
\begin{equation}
d=2\cos[\pi/(k+2)] \,.
\label{E15}
\end{equation}
In particular for the special value $k=3$, we have the golden ratio $d=(1+\sqrt 5)/2$ which is the solution of the algebraic equation $d^{2}=d+1$. 
For $k=2$, the Baraha number $d$ and the quantum dimension 
$d_{2j=1}=\sqrt{2}$ coincide. 

To clarify the meaning of the $d$'s one should emphasize that
(i) the numbers $d$ determine eigen values of Wilson operator for the contractible unknotted loop \cite{FF,FNSWW}. (ii) For special values of the parameter $d=q+q^{-1}$, the generators  $B_{i}=I-qe_{i}$ satisfy the relations
of a braid group under the
condition, that the generators $e_{i}$ satisfy the relation $e_{i}^{2}=de_{i}$ of the Temperley-Lieb algebra.
(iii) The values of
the parameter $d$ are nontrivial restriction, which leads to the
finite-dimensional Hilbert spaces.
(iv) The wavefunction $\Psi$, defined on the
one-dimensional manifold, which is a joining up of the
arbitrary tangle $\alpha$ and the Wilson loop $\bigcirc$,
i.e. $\Psi(\alpha \cup \bigcirc)$, equals $d\Psi(\alpha)$ \cite{FNSWW}.
Thus the parameter $d$ has the meaning
of the weight of the contractible unknotted Wilson loop, and
$d^{2}$ acquires the meaning of fugacity \cite{FFNWW}.
(v) Besides, it turns out, that for the
mentioned values of $d$, the theory is unitary.

Summarizing one can say that the points of intersection of braid statistics and statistics with the generalized exclusion principle are  the set of the Baraha points $d=2\cos[\pi/(k+2)]$, where the processes of braiding and fusion of string manifolds are self-consistently
united.

\section{Discussion} 

The problem of the coexistence \cite{AFF} of locality and braiding 
can be solved by constructing Hamiltonians $H=\sum_{i}H_{i}$ of the Rokhsar-Kivelson (RK) 
type \cite{RokKiv}. Each term $H_{i}=Q^{+}_{i}Q_{i}$  
in the sum with  
$Q =\left( \begin{array}{cc}
 1  & -1  \\
 -1 & 1
\end{array} \right)$ acts at the RK point as a projector builted by dimer configurations.  If we locate the dimers on the opposite links of plaquetts we will encounter contradiction due
to spatial separation of the braiding phenomena. 
The best way to solve this problem is the distribution 
of dimer states between odd and even sites of the 
lattice \cite{FTLTKWF}. Another way corresponds to the use of the Fischer lattice \cite{Fish}. The checkerboard distribution of linked dimer 
degrees of freedom with defects in the order of effective fluxes (Fig. 2) may be one of the possible ways to solve the problem.
\begin{figure}
\begin{center}
\includegraphics[width=0.7\linewidth]{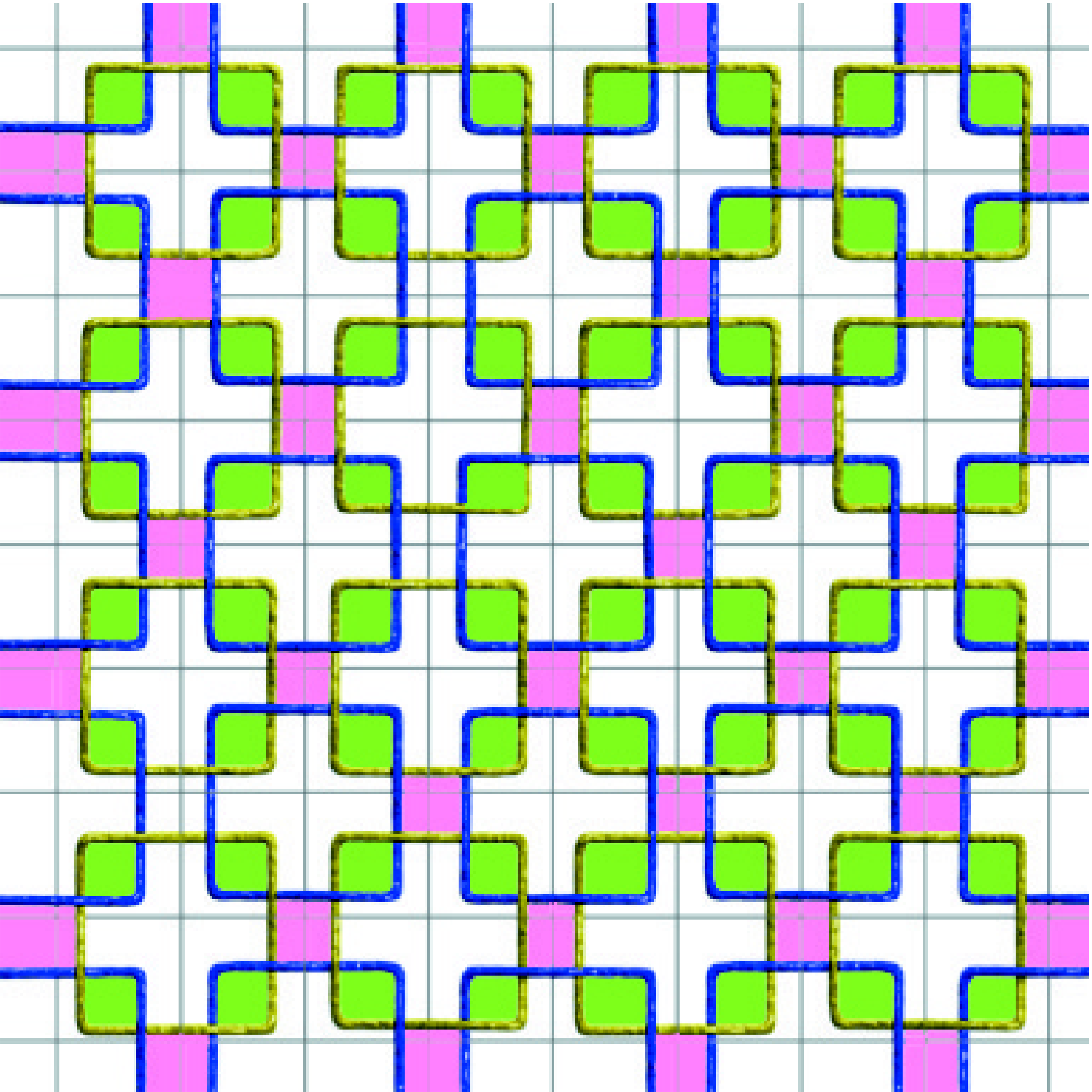}
\end{center}
\caption{(Color online). 
A planar representation of the net.  
Shaded regions show distributions of the fluxes.  
}
\end{figure}

The realistic candidate for the Hamiltonian with such a type of the ground state is the Hamiltonian $H_{i}$ which contains the Temperley-Lieb generators $e_{i}$ 
in the form of the projectors 
\begin{equation}
H_{i}=\frac{1}{d}e_{i}
\label{G10}
\end{equation}
to the singlet states. Obviously $H_{i}^{2}=H_{i}$ due to the Temperley-Lieb commutation 
relation $e_{i}^{2}=de_{i}$. 
The $d$'s here are the Baraha numbers from the second section. 
Because of the rank-level symmetry, 
i.e. $SU(N)_{k}=SU(k)_{N}$, and the argument based on small 
values of integers, 
the $SU(2)_{2}\times {\overline S\overline U(2)}_{2}$ theory is a good candidate. The matrix of the $6j$-symbols \cite{P} in this case is equal to $\pm \frac{1}{\sqrt 2}\left( \begin{array}{cc}
 1 & 1  \\
 1 & -1
\end{array} \right)$  and the braid operator is 
$\left( \begin{array}{cc}
 1 & 0  \\
 0 & i
\end{array} \right)$. 
Another important model with computational universal rules is based \cite{FTLTKWF} on a golden chain builted by $SU(2)_{3}\times \overline S \overline U(2)_{3}$ Fibonacci anyons. To construct $2D$ nets consisting of Fibonacci golden chains we should employ the Kirby calculus. This calculus is based on the application of Hopf links and is widely used in the theory of 3-manifolds. 

In summary, we found the quantum dimensions as exact solutions of discrete 
equations encoding brading and fusion processes. By means of projection of 
knotted field configurations we proposed  the quantum dimer models which 
incorporated braiding properties of one-dimensional topology.    

\section*{Acknowledgments}

This work was supported 
in part by the E.I.N.S.T.E.IN grant (M.L., A.P.), RFBR grants (A.P., V.V.) 
Nos. 06-02-16561, 06-02-92052, and the programs (A.P.) RNP 2.1.1 (grant No. 2369) 
and "Problems of nonlinear dynamics" of Presidium of the Russian Academy 
of Sciences. The authors aknowledge also the INFN for support in part by the project LE41. 

\section{Appendix} 

Let us show that the Eq. (\ref{E7}) is a particular case of the Hirota equation \cite{KLWZ} 
\begin{equation}
T^{a}_{t}(u+1)T^{a}_{t}(u-1)-T^{a}_{t+1}(u)T^{a}_{t-1}(u)=
T^{a+1}_{t}(u)T^{a-1}_{t}(u) \, .
\label{E9}
\end{equation}
Here $a$ is the index of the agebra $A_{k+1}$, $t$ is the discrete time
and $u$ is the discrete values of rapidities. The functions $T_{t}^{a}(u)$ are the eigen values of the transfer-matrix \cite{KLWZ}.
For the gauged functions
$Y^{a}_{t}(u)=T^{a}_{t+1}(u)T^{a}_{t-1}(u)/\left(T^{a+1}_{t}(u)
T^{a-1}_{t}(u)\right)$  the following equation
\begin{equation}
Y^{a}_{t}(u+1)Y^{a}_{t}(u-1)=
\frac{(1+Y^{a}_{t+1}(u))(1+Y^{a}_{t-1}(u))}{(1+(Y^{a+1}_{t}(u))^{-1})
(1+(T^{a-1}_{t}(u))^{-1})}
\label{E10} 
\end{equation}
is valid. In the $A_{1}$-algebra case the function $Y^{1}_{t}(u) \equiv Y_{t}(u)$ satisfies the equation 
\begin{equation}
Y_{t}(u+1)Y_{t}(u-1)=
(1+Y_{t+1}(u))(1+Y_{t-1}(u)). 
\label{E11} 
\end{equation}
Putting $Y_{t}=b_{t}^{2}-1$ in this equation, in the limit $u \gg 1$ we get the equation $b_{t}^{2}=1+b_{t+1}b_{t-1}$ which coincides with (\ref{E7}). We see
that the function $Y_{t}$ equals $w_{t}$.

\end{document}